\documentclass[english]{article}
\usepackage[T1]{fontenc}
\usepackage[latin9]{inputenc}
\usepackage{amsmath}
\usepackage{amssymb}
\usepackage{esint}
\usepackage{babel}
\begin{document}

\title{Matrix mechanics of the relativistic point particle and string in
Clifford space}

\author{Kaare Borchsenius}
\maketitle
\begin{abstract}
We resolve the space-time canonical variables of the relativistic
point particle into inner products of Weyl spinors with components
in a \mbox{Clifford} algebra and find that these spinors themselves
form a canonical system with generalized Poisson brackets. For $N$
particles, the inner products of their Clifford coordinates and momenta
form two $N\times N$ Hermitian matrices $X$ and $P$ which transform
under a $U(N)$ symmetry in the generating algebra. This is used as
a starting point for deriving matrix mechanics for a point particle
in Clifford space. Next we consider the string. The Lorentz metric
induces a metric and a scalar on the world sheet which we represent
by a Jackiw-Teitelboim term in the action. The string is described
by a polymomenta canonical system and we find the wave solutions to
the classical equations of motion for a flat world sheet. Finally,
we show that the $SL(2.\mathbb{C})$ charge and space-time momentum
of the quantized string satisfy the Poincaré algebra. 
\end{abstract}

\section{Introduction}

It is well known that a null vector can be resolved into a product
of two Weyl spinors 
\[
x^{A\dot{B}}=c^{A}\cdot c^{*\dot{B}},\quad x^{\mu}x_{\mu}=0,
\]
 where $x^{A\dot{B}}$ and $x^{\mu}$ are related through the equivalence
between real four-vectors and second-rank hermitian spinors 
\[
V^{\mu}=\frac{1}{2}\sigma_{A\dot{B}}^{\mu}V^{A\dot{B}},\quad V^{A\dot{B}}=\sigma_{\mu}^{A\dot{B}}V^{\mu},
\]
 and $\sigma_{\mu}$ are the four hermitian matrices which extend
the Pauli matrices \cite{key-17}. To resolve non-null vectors, we
need something like 
\begin{equation}
x^{A\dot{B}}=c^{A}\bullet c^{*\dot{B}},\label{eq:b3}
\end{equation}
 where $\bullet$ is a product which belongs to some non-commutative
algebra. This problem can be compared to the somewhat similar problem
of resolving the Lorentz metric $\eta_{\mu\nu}$ into vectors. The
well known solution is $\eta_{\mu\nu}=\frac{1}{2}\{\gamma_{\mu},\gamma_{\nu}\}$
where the Dirac matrices $\gamma_{\mu}$ generate the Clifford algebra
$\mathcal{C}l(1,3,\mathbb{R}$). The components of any real symmetric
$4\times4$ matrix of signature $(1,3)$ can therefore be expressed
as the inner products (anti-commutators) of vectors (real linear combinations
of $\gamma$ matrices) belonging to $\mathcal{C}l(1,3,\mathbb{R})$.
Real Clifford algebras are associated with real quadratic forms, but
there is no similar relationship between hermitian sesquilinear forms
and complex Clifford algebras $\mathcal{C}l(\mathbb{C})$. Instead
we must use even-dimensional real Clifford algebras written in complex
form \cite{key-1,key-3,key-4,key-5} Consider a future directed time-like
vector $x^{\mu}$. A unitary transformation followed by a non-uniform
scaling can reduce $x^{A\dot{B}}$ to a diagonal matrix with ones
in the diagonal and can be effected by a suitable linear transformation
of $c^{A}$ so that (\ref{eq:b3}) becomes 
\[
c_{i}\bullet c_{j}^{*}=\delta_{ij}.
\]
 This can be compared to the algebra of creation and annihilation
operators for two fermions 
\begin{equation}
\{\mathbf{a}_{i},\mathbf{a}_{i}^{\dagger}\}=\delta_{ij}\cdot1,\quad\{\mathbf{a}_{i},\mathbf{a}_{j}\}=0,\quad i,j=1,2.\label{eq:b5}
\end{equation}
 Defining $e_{i}=\mathbf{\mathit{i}(a}_{i}+\mathbf{a}_{i}^{\dagger})\,,\,\,\,e_{2+i}=\mathbf{a}_{i}-\mathbf{a}_{i}^{\dagger}\,,\,\,i=1,2\:$
, the anti-commutation relations (\ref{eq:b5}) become 
\[
\{e_{i},e_{j}\}=-2\delta_{ij},\quad i,j=1,\ldots,4
\]
 which generate the Clifford algebra $\mathcal{C}l(0,4,\mathbb{R})$.
This suggests that a solution to (\ref{eq:b3}) would be to use spinors
with values in the split Clifford algebra $\mathcal{C}l(4,4,\mathbb{R})$
and to let $\bullet$ be the inner product (anti-commutator) of this
algebra. This expectation is borne out by the following proposition 
\begin{quote}
\emph{Let} $V_{C}$ \emph{be a} $2n$\emph{\nobreakdash-dimensional
complex linear space with complex conjugation} {*} \emph{and} $H$
\emph{an} $n\times n$\emph{ Hermitian matrix of arbitrary signature.
Then the components of} $H$ \emph{can be expressed as} 
\[
H_{ij}=c_{i}\bullet c_{j}^{*},\quad,c_{i}\bullet c_{j}=0,\quad i,j=1,\ldots,n
\]
 \emph{where} $c_{i}$ \emph{belong to} $V_{C}$ \emph{and} $\bullet$
\emph{is the inner product} 
\[
a\bullet b\equiv\frac{1}{2}\{a,b\}
\]
 \emph{of the Clifford algebra} $\mathcal{C}l(2n,2n,\mathbb{R})$
\emph{on the} $4n$ \emph{dimensional real linear space} $V_{R}$
\emph{which corresponds to }$V_{C}$ . 
\end{quote}
Proof. Let $e_{i},\,\,f_{i},\,\,i=1,\ldots,n$ be a basis for $V_{C}$
and $g_{i}=i(e_{i}+e_{i}^{*}),\:g_{n+i}=e_{i}-e_{i}^{*},\:h_{i}=i(f_{i}+f_{i}^{*}),\:h_{n+i}=f_{i}-f_{i}^{*},\:i=1,\ldots n$
a basis for $V_{R}$ . Let $g_{i}$ and $h_{i}$ generate the Clifford
algebra $\mathcal{C}l(2n,2n,\mathbb{R})$ on $V_{R}$ through 
\[
g_{i}\bullet g_{j}=2\delta_{ij},\quad h_{i}\bullet h_{j}=-2\delta_{ij},\quad g_{i}\bullet h_{j}=0,\quad i,j=1,\ldots,2n.
\]
 Then the basis $e_{i},\,f_{i}$ for $V_{C}$ satisfies 
\[
e_{i}\bullet e_{j}^{*}=-\delta_{ij},\quad f_{i}\bullet f_{j}^{*}=\delta_{ij},\quad e_{i}\bullet e_{j}=f_{i}\bullet f_{j}=0,\quad i,j=1,\ldots,n.
\]
 We can create any $n\times n$ diagonal matrix of plus or minus ones
by setting $c_{i}$ equal to either $f_{i}$ or $e_{i}$ . A zero
in the $k$\nobreakdash-th entry of the diagonal can be created by
$c_{k}=e_{k}+f_{k}$. A non-uniform scaling followed by a unitary
transformation can transform this diagonal matrix into any desired
$n\times n$ hermitian matrix with the same signature and can be effected
by a suitable complex linear transformation of the $c$'s $\Box$

We shall resolve both the coordinates and momenta of the point particle
into Clifford spinors 
\begin{equation}
x^{A\dot{B}}=c^{A}\bullet c^{*\dot{B}},\quad p_{A\dot{B}}=d_{A}^{*}\bullet d_{\dot{B}},\label{eq:b11}
\end{equation}
 but we also need the Clifford algebra to be large enough that the
inner products $c^{A}\bullet d_{B}^{*}$ are algebraically independent
of $x$ and $p$. This can be accomplished, for example by enlarging
$\mathcal{C}l(4,4,\mathbb{R})$ to $\mathcal{C}l(8,8,\mathbb{R})$
and then generating $x$ and $p$ by each their own $\mathcal{C}l(4,4,\mathbb{R})$
subalgebras. This makes $c\bullet d^{*}$ vanish. The second step
is to choose two Clifford elements $h_{i}$ whose inner products with
both $c$ and $d^{*}$ vanish, and to make the substitution 
\[
c^{A}\rightarrow c^{A}+A_{i}^{A}h_{i},\quad d_{A}^{*}\rightarrow d_{A}^{*}+B_{iA}h_{i}^{*}.
\]
 This will change $x$ and $p$ only by additive matrices that will
not constrain them, and the two matrices $A$ and $B$ can be adjusted
to produce any desired value of $c\bullet d^{*}$. Apart from this
requirement, the dimension of the single-particle Clifford algebra
is not of importance in our discussion.

Note that $c^{*\dot{A}}$ and $d_{A}^{*}$ have the same commutation
properties but transform differently under $SL(2.\mathbb{C})$. The
complex conjugation symbol $*$ can therefore not be omitted, as it
often is, because it specifies the commutation properties of the element
in question. It is tacitly assumed that the inner product of elements
of the same kind vanishes, and this will not be written out explicitly.

If the variation of a function $f$ of $c$ with respect to a variation
of $c$ can be expressed on the form 
\[
\delta f=\frac{1}{2}\{\frac{\partial f}{\partial c^{A}},\delta c^{A}\},
\]
 we shall call $\partial f/\partial c^{A}$ a derivative of $f$ with
respect to $c$. It is defined up to terms which anti-commute with
arbitrary variations $\delta c$. Trivially, we have $\partial c^{A}/\partial c^{B}=\delta_{B}^{A}.$
From (\ref{eq:b11}) it follows that a differentiable function $f$
of $x^{\mu}$ has the derivative 
\begin{equation}
\frac{\partial f(x)}{\partial c^{A}}=\frac{\partial f(x)}{\partial x^{\mu}}\frac{1}{2}\sigma_{A\dot{B}}^{\mu}c^{*\dot{B}}=\frac{\partial f(x)}{\partial x^{A\dot{B}}}c^{*\dot{B}}.\label{eq:b15}
\end{equation}

\section{Clifford substructure of the relativistic point particle}

Let the space-time coordinates and momenta of the relativistic point
particle be resolved into Clifford spinors according to (\ref{eq:b11}).
The equations of motion are obtained from the condition that the re\-para\-metri\-zation
invariant action 
\begin{equation}
I=4\sqrt{m}\int d\tau\sqrt[4]{\frac{1}{2}\frac{dc^{A}}{d\tau}\bullet\frac{dc^{*\dot{B}}}{d\tau}\,\frac{dc_{A}}{d\tau}\bullet\frac{dc_{\dot{B}}^{*}}{d\tau}}\label{eq:c2}
\end{equation}
 is stationary under arbitrary variations of $c(\tau)$. The momenta
conjugate to $c$ are 
\[
d_{A}^{*}\equiv\frac{\partial L}{\partial\frac{dc^{A}}{d\tau}}=\sqrt{m}\,\bigl(\frac{1}{2}\frac{dc^{E}}{d\tau}\bullet\frac{dc^{*\dot{F}}}{d\tau}\;\frac{dc_{E}}{d\tau}\bullet\frac{dc_{\dot{F}}^{*}}{d\tau}\bigl)^{-\frac{3}{4}}\bigl(\frac{dc_{A}}{d\tau}\bullet\frac{dc_{\dot{B}}^{*}}{d\tau}\bigl)\,\frac{dc^{*\dot{B}}}{d\tau},
\]
 and, as expected, the Hamiltonian vanishes. A straightforward calculation
using the four-vector identity 
\begin{equation}
V_{A\dot{E}}V^{B\dot{E}}=\frac{1}{2}\delta_{A}^{B}V_{F\dot{E}}V^{F\dot{E}}\label{eq:c30}
\end{equation}
 shows that the conjugate momenta $d_{A}^{*}$ satisfy the constraint
\begin{equation}
p^{\mu}p_{\mu}-m^{2}=0,\label{eq:c4}
\end{equation}
 where $p_{\mu}$ are the space-time momenta defined in (\ref{eq:b11}).
This happens to be the same constraint as would have been obtained
from the space-time action $\int d\tau\,\sqrt{\dot{x}^{2}}$. According
to constrained dynamics, the Hamiltonian density is proportional to
the constraint 
\begin{equation}
\mathcal{H}(p,e(\tau))=e(\tau)(p{}^{\mu}p_{\mu}-m^{2}),\label{eq:c5}
\end{equation}
 where $e(\tau)$ is an einbein. This Hamiltonian can also be obtained
from the Polyakov (`metrical') type of action 
\begin{equation}
I=\int d\tau\Bigl(3e(\tau)^{-\frac{1}{3}}\sqrt[3]{\frac{1}{2}\frac{dc^{A}}{d\tau}\bullet\frac{dc^{*\dot{B}}}{d\tau}\,\frac{dc_{A}}{d\tau}\bullet\frac{dc_{\dot{B}}^{*}}{d\tau}}+m^{2}\,e(\tau)\Bigr),\label{eq:c21}
\end{equation}
 which recovers (\ref{eq:c2}) when the equations of motion for the
einbein $e(\tau)$ are substituted back into the action. The momenta
conjugate to $c$ are 
\[
d_{A}^{*}=e(\tau)^{-\frac{1}{3}}\bigl(\frac{1}{2}\frac{dc^{E}}{d\tau}\bullet\frac{dc^{*\dot{F}}}{d\tau}\,\frac{dc_{E}}{d\tau}\bullet\frac{dc_{\dot{F}}^{*}}{d\tau}\bigr)^{-\frac{2}{3}}\bigl(\frac{dc_{A}}{d\tau}\bullet\frac{dc_{\dot{B}}^{*}}{d\tau}\bigr)\,\frac{dc^{*\dot{B}}}{d\tau},
\]
 which determine the Hamiltonian density 
\begin{equation}
\mathcal{H}(c,d)=d_{A}^{*}\bullet\frac{dc^{A}}{d\tau}+c.c.-L,\label{eq:c24}
\end{equation}
 where $c.c.$ denotes the complex conjugate of the previous term
and $L$ is the Lagrangian in (\ref{eq:c21}). A straightforward calculation
gives 
\[
d_{A}^{*}\bullet\frac{dc^{A}}{d\tau}+c.c.=4e(\tau)^{-\frac{1}{3}}\sqrt[3]{\frac{1}{2}\frac{dc^{A}}{d\tau}\bullet\frac{dc^{*\dot{B}}}{d\tau}\,\frac{dc_{A}}{d\tau}\bullet\frac{dc_{\dot{B}}^{*}}{d\tau}}\,,
\]
 
\[
p^{\mu}p_{\mu}\equiv\frac{1}{2}d{}^{A}\bullet d{}^{*\dot{B}}\,d{}_{A}\bullet d{}_{\dot{B}}^{*}=e(\tau)^{-\frac{4}{3}}\sqrt[3]{\frac{1}{2}\frac{dc^{A}}{d\tau}\bullet\frac{dc^{*\dot{B}}}{d\tau}\,\frac{dc_{A}}{d\tau}\bullet\frac{dc_{\dot{B}}^{*}}{d\tau}}\,,
\]
 which, when applied to (\ref{eq:c24}), gives the Hamiltonian (\ref{eq:c5})
of constrained dynamics. Hence the first order (Hamiltonian) form
of the action (\ref{eq:c2}) is 
\begin{equation}
I=\int d\tau\Bigl(d_{A}^{*}\bullet\frac{dc^{A}}{d\tau}+c.c.-e(\tau)(p{}^{\mu}p_{\mu}-m^{2})\Bigr).\label{eq:c6}
\end{equation}
 This action has a global $SL(2.\mathbb{C})$ and $U(1)$ gauge symmetry
with the conserved Noether charges 
\[
\mathcal{J_{AB}}\equiv d_{A}^{*}\bullet c_{B}+d_{B}^{*}\bullet c_{A},\quad\jmath\equiv i(d_{A}^{*}\bullet c^{A}-d_{\dot{A}}\bullet c^{*\dot{A}}).
\]
 To obtain the correct space-time equations of motion, it is necessary
to assume (as an initial value condition) that they vanish 
\begin{equation}
d_{A}^{*}\bullet c_{B}+d_{B}^{*}\bullet c_{A}=0,\label{eq:c10}
\end{equation}
 
\begin{equation}
d_{A}^{*}\bullet c^{A}-d_{\dot{A}}\bullet c^{*\dot{A}}=0.\label{eq:c11}
\end{equation}
 Since all skew-symmetric second rank spinors are proportional to
$\epsilon_{AB}$, (\ref{eq:c10}) gives 
\begin{equation}
d_{A}^{*}\bullet c_{B}=\mu(\tau)\,\epsilon_{AB},\quad\text{or}\quad d_{A}^{*}\bullet c^{B}=\mu\delta_{A}^{B},\label{eq:c12}
\end{equation}
 with (\ref{eq:c11}) saying that $\mu(\tau)$ is real. For short,
we shall refer to this condition as the `Noether constraint'. The
canonical equations of motion are obtained by independent variation
of $c$ and $d$ 
\begin{equation}
\frac{dc^{A}}{d\tau}=\frac{\partial\mathcal{H}}{\partial d_{A}^{*}}=\frac{\partial\mathcal{H}}{\partial p_{A\dot{E}}}d_{\dot{E}},\quad\frac{d\,d_{A}^{*}}{d\tau}=-\frac{\partial\mathcal{H}}{\partial c^{A}}=-\frac{\partial\mathcal{H}}{\partial x^{A\dot{E}}}c^{*\dot{E}},\label{eq:c7}
\end{equation}
 where we have used the differentiation rule (\ref{eq:b15}). Taking
the inner product of these equations with $c^{*\dot{B}}$ and $d$$_{\dot{B}}$
gives 
\[
\frac{dx^{A\dot{B}}}{d\tau}=2\frac{\partial\mathcal{H}}{\partial p_{A\dot{E}}}\:c^{*\dot{B}}\bullet d_{\dot{E}},\quad\frac{dp_{A\dot{B}}}{d\tau}=-2\frac{\partial\mathcal{H}}{\partial x^{A\dot{E}}}\:c^{*\dot{E}}\bullet d_{\dot{B}},
\]
 which by use of the Noether constraint (\ref{eq:c12}) become 
\[
\frac{dx^{A\dot{B}}}{d\tau}=2\frac{\partial\mathcal{H}}{\partial p_{A\dot{B}}}\mu(\tau),\quad\frac{dp_{A\dot{B}}}{d\tau}=-2\frac{\partial\mathcal{H}}{\partial x^{A\dot{B}}}\,\mu(\tau).
\]
 In the parametrization 
\begin{equation}
\overline{e}(\overline{\tau})=\frac{1}{2m\mu(\bar{\tau})},\label{eq:c17}
\end{equation}
 these equations reduce to the canonical equations of motion 
\begin{equation}
\frac{dx^{\mu}}{d\overline{\tau}}=\frac{\partial\mathcal{H}}{\partial p_{\mu}},\quad\frac{dp_{\mu}}{d\overline{\tau}}=-\frac{\partial\mathcal{H}}{\partial x^{\mu}},\quad\mathcal{H}(x,p)\equiv\frac{1}{2m}(p^{\mu}p_{\mu}-m^{2}),\label{eq:c18}
\end{equation}
 for a relativistic point particle with proper time $\overline{\tau}$.
This proper time is not defined at points where $\mu$ vanishes. There
will be just one such point and it represents a `turning point' where
the space-time trajectory has an endpoint and the underlying trajectory
in Clifford space starts to reproduce it for the second time. From
(\ref{eq:c7}) and the Hamiltonian constraint (\ref{eq:c4}), we obtain
an explicit expression for $\mu(\tau)$ 
\[
\frac{d}{d\tau}\mu(\tau)=\frac{d}{d\tau}(\frac{1}{2}d_{E}^{*}\bullet c^{E})=e(\tau)m^{2},\quad\mu(\tau)=\int_{\tau_{0}}^{\tau}dt\,m^{2}e(t).
\]
 Hence $\mu(\tau)$ is determined by the mass of the particle and
the `turning point' $\tau_{0}$ of its motion.

The fact that the Noether constraint (\ref{eq:c12}) leads to the
conventional equations of motion (\ref{eq:c18}) for $x$ and $p$,
can be understood in terms of generalized Poisson brackets. We define
the Poisson bracket in Clifford space as the `Clifford bracket' 
\begin{multline*}
\left\{ N,M\right\} _{C.B.}\equiv\frac{1}{2}\Bigl(\{\frac{\partial N}{\partial c^{A}},\frac{\partial M}{\partial d_{A}^{*}}\}+\{\frac{\partial N}{\partial c^{*\dot{A}}},\frac{\partial M}{\partial d_{\dot{A}}}\}-\{\frac{\partial M}{\partial c^{A}},\frac{\partial N}{\partial d_{A}^{*}}\}-\{\frac{\partial M}{\partial c^{*\dot{A}}},\frac{\partial N}{\partial d_{\dot{A}}}\}\Bigr),
\end{multline*}
 where $\{,\}$ denotes the anti-commutator. This bracket is skew-symmetric
in $N$ and $M$ and real when $N$ and $M$ are real. The equations
of motion (\ref{eq:c7}) can be written in terms of brackets 
\[
\frac{dc^{A}}{d\tau}=\left\{ c^{A},\mathcal{H}\right\} _{C.B.},\quad\frac{d\,d_{A}^{*}}{d\tau}=\left\{ d_{A}^{*},\mathcal{H}\right\} _{C.B.},
\]
 which leads to 
\begin{equation}
\frac{dx^{\mu}}{d\tau}=\left\{ x^{\mu},\mathcal{H}\right\} _{C.B.},\quad\frac{d\,p_{\mu}}{d\tau}=\left\{ p_{\mu},\mathcal{H}\right\} _{C.B.}.\label{eq:c31}
\end{equation}
 In general, these equations cannot be expressed solely in terms of
$x$ and $p$. However, when the Noether constraint (\ref{eq:c12})
holds, then by use of the differentiation rule (\ref{eq:b15}), the
Clifford bracket becomes proportional to the ordinary Poisson bracket:
\begin{multline}
\left\{ N(x,p),M(x,p)\right\} _{C.B.}=\Bigl(\frac{\partial N}{\partial x^{\mu}}\frac{\partial M}{\partial p_{\nu}}-\frac{\partial M}{\partial x^{\mu}}\frac{\partial N}{\partial p_{\nu}}\Bigr)\Bigl(\frac{1}{8}\sigma_{A\dot{B}}^{\mu}\sigma_{\nu}^{A\dot{F}}\{c^{*\dot{B}},d_{\dot{F}}\}+c.c.\Bigr)\\
=\mu\Bigl(\frac{\partial N}{\partial x^{\mu}}\frac{\partial M}{\partial p_{\mu}}-\frac{\partial M}{\partial x^{\mu}}\frac{\partial N}{\partial p_{\mu}}\Bigr)=\mu\left\{ N(x,p),M(x,p)\right\} _{P.B.}.\label{eq:c23}
\end{multline}
 After a reparametrization which absorbs $\mu$, (\ref{eq:c23}) turns
the equations of motion (\ref{eq:c31}) into the usual space-time
form (\ref{eq:c18}) which `hides' the Clifford substructure.

\section{System of N particles with a U(N) symmetry}

Assuming that the Clifford algebra for the point particle is $Cl(2n,2n,\mathbb{R})$,
we can accommodate $N$ particles in $Cl(2nN,2nN,\mathbb{R})$ in
such a way that all inner products between Clifford coordinates and
momenta belonging to different particles vanish. The generating algebra
\begin{equation}
e_{i}^{p}\bullet e_{j}^{*q}=\delta_{ij}\delta_{pq}\,sign(p),\quad e_{i}^{p}\bullet e_{j}^{q}=0,\quad i,j=1,\ldots N,\quad p,q=1,\ldots2n,\label{eq:d20}
\end{equation}
 where $sign(p)$ denotes the sign of $e^{p}\bullet e^{*p}$, is preserved
by the $U(N)$ unitary transformation 
\[
e_{i}^{p}\rightarrow U_{ih}\,e_{h}^{p},\quad U_{ih}U_{jh}^{*}=\delta_{ij}.
\]
 If we assemble the canonical variables $c_{i}^{A}$ and $d_{iA}^{*},\,i=1\ldots,N$
of the $N$ particles into the ket- and bra-vectors $\stackrel{>}{C^{A}}$
and $\stackrel{<}{D_{A}}$ respectively, then the corresponding space-time
coordinates and momenta are elements of the $N\times N$ diagonal
matrices 
\[
X^{A\dot{B}}=\stackrel{>}{C^{A}}\bullet\stackrel{<\dot{B}}{C\quad},\quad P_{A\dot{B}}=\stackrel{>}{D_{\dot{B}}}\bullet\stackrel{<}{D_{A}},
\]
 which trivially satisfy the commutation relations 
\[
[X^{\mu},X^{\nu}]=[P_{\mu},P_{\nu}]=[X^{\mu},P_{\nu}]=0.
\]
 The equations of motion for this dynamical system can be derived
from the sum 
\begin{equation}
I=\int d\tau\,Tr\,\Bigl(\frac{d}{d\tau}\stackrel{>}{C^{A}}\bullet\stackrel{<}{D_{A}}+c.c.-\mathcal{H}\,\Bigr),\quad\mathcal{H}\equiv e(\tau)\bigl(P^{\mu}P_{\mu}-m^{2}\cdot\underline{1}\bigr),\label{eq:d6}
\end{equation}
 of the single-particle actions (\ref{eq:c6}). The Noether constraint
(\ref{eq:c12}) becomes 
\begin{equation}
\stackrel{>}{C^{A}}\bullet\stackrel{<}{D_{B}}=\mu(\tau)\,\delta_{B}^{A}\cdot\underline{1}.\label{eq:d7}
\end{equation}
 We observe that (\ref{eq:d6}) and (\ref{eq:d7}) are preserved by
the global $U(N)$ transformations 
\begin{equation}
\stackrel{>}{C^{A}}\rightarrow U\stackrel{>}{\,C^{A},}\quad\stackrel{<}{D_{A}}\rightarrow\stackrel{<}{D_{A}\,}U^{\dagger},\label{eq:d9}
\end{equation}
 which produce the similarity transformations 
\[
X^{\mu}\rightarrow UX^{\mu}U^{\dagger},\quad P_{\mu}\rightarrow UP_{\mu}U^{\dagger}
\]
 of the Hermitian matrices $X^{\mu}$ and $P_{\mu}$. Such transformations
create off-diagonal entries in $X$ and $P$ which correspond to artificial
couplings between the Clifford coordinates and momenta belonging to
different particles.

The motion of a classical point particle can be described by a set
of integral curves in the phase space $(x^{\mu},p_{\mu})$. From the
foregoing it follows that the coordinates of these integral curves
are eigenvalues of $X^{\mu}$ and $P_{\mu}$.

\section{Matrix Mechanics}

The unitary system obtained in the foregoing has the structure of
a finite dimensional form of matrix mechanics and suggests the quantization
\begin{multline}
c\rightarrow\stackrel{>}{C},\quad d^{*}\rightarrow\stackrel{<}{D},\quad c\bullet d^{*}\rightarrow\stackrel{>}{C}\bullet\stackrel{<}{D},\quad c\bullet c^{*}\rightarrow\stackrel{>}{C}\bullet\stackrel{<}{C},\quad d\bullet d^{*}\rightarrow\stackrel{>}{D}\bullet\stackrel{<}{D},\label{eq:e20}
\end{multline}
 
\begin{multline}
\left\{ c^{A},M(x,p)\right\} _{C.B.}\rightarrow\frac{1}{i\hslash}[X^{A\dot{B}},M(X,P)]\stackrel{>}{D_{\dot{B}}},\\
\left\{ d_{A}^{*},M(x,p)\right\} _{C.B.}\rightarrow\frac{1}{i\hslash}\stackrel{<\dot{B}}{C\quad}[P_{A\dot{B}},M(X,P)].\label{eq:e21}
\end{multline}
 By use of the Noether constraint (\ref{eq:d7}), it follows that
\[
\left\{ N(x,p),M(x,p)\right\} _{C.B.}\rightarrow\frac{\mu}{i\hslash}\left[N(X,P),M(X,P)\right],
\]
 and consequently the well-known rule $\{,\}_{P.B.}\rightarrow\frac{1}{i\hslash}[,]$.
To show that this is a valid procedure, we shall derive matrix mechanics
from a variational principle. If we simply used the sum of the single-particle
actions, the Noether constraint would become too weak. Instead, we
must require that all real linear combinations of the single-particle
actions are stationary and obey the constraints. This corresponds
to the action 
\begin{equation}
I=\int d\tau\sum_{i=1}^{N}\phi_{i}L_{i},\quad L_{i}=d_{iA}^{*}\bullet\frac{dc_{i}^{A}}{d\tau}+c.c.-\mathcal{H}(p_{i},e(\tau)),\label{eq:e1}
\end{equation}
 where the coefficients $\phi_{i}$ are arbitrary real constants,
and $L_{i}$ are the single-particle Lagrangians. When $\Phi$ denotes
the $N\times N$ diagonal matrix with $\phi_{i}$ along its diagonal
and $P$ is diagonal, the action (\ref{eq:e1}) can be written as
\begin{equation}
I=\int d\tau Tr\,\biggl(\,\Phi\bigl(\,\frac{d\stackrel{>}{C^{A}}}{d\tau}\bullet\stackrel{<}{D_{A}}+h.c.-\mathcal{H}\bigr)\biggr),\quad\mathcal{H}\equiv e(\tau)(P^{\mu}P_{\mu}-m^{2}\cdot\underline{1}).\label{eq:e6}
\end{equation}
 This action is preserved by the unitary transformation (\ref{eq:d9})
with $\Phi$ transforming according to 
\[
\overline{\Phi}\rightarrow U\Phi U^{\dagger},\quad\Phi^{\dagger}=\Phi.
\]
 The diagonal matrices $\Phi$ and $P_{\mu}$ trivially satisfy the
unitarily invariant conditions 
\begin{equation}
\frac{d\Phi}{d\tau}=[\Phi,P_{\mu}]=[P_{\mu},P_{\nu}]=0.\label{eq:e0}
\end{equation}
 Conversely, these conditions ensure that the action (\ref{eq:e6})
can be gauged back into (\ref{eq:e1}). The conserved $SL(2.\mathbb{C})$
and $U(N)$ Noether charges corresponding to the action (\ref{eq:e6})
are 
\[
J_{AB}=Tr\Bigl(\bigl(\Phi\stackrel{>}{(C_{A}}\bullet\stackrel{<}{D_{B}}+\stackrel{>}{C_{B}}\bullet\stackrel{<}{D_{A}})\Bigr),\qquad j=i(\Phi\stackrel{>}{C^{A}}\bullet\stackrel{<}{D_{A}}-h.c.).
\]
 Requiring that they vanish for all values of $\Phi$ gives (\ref{eq:d7}).

For Hamiltonians which are polynomial functions of $X$ and $P$,
the derivatives of $Tr(\mathcal{H})$ with respect to $\stackrel{>}{C}$
and $\stackrel{<}{D}$ are well defined and can be written on the
form 
\[
\frac{\partial Tr(\mathcal{H})}{\partial\stackrel{>}{C^{A}}}=\stackrel{<\dot{E}}{C\quad}\frac{\partial Tr(\mathcal{H})}{\partial X^{A\dot{E}}},\quad\frac{\partial Tr(\mathcal{H})}{\partial\stackrel{<}{D_{A}}}=\frac{\partial Tr(\mathcal{H})}{\partial P_{A\dot{E}}}\stackrel{>}{D_{\dot{E}}},
\]
where $\partial Tr(\mathcal{H})/\partial X$ and $\partial Tr(\mathcal{H})/\partial P$
are matrix functions of $X$ and $P$. The equations of motion are
obtained by requiring the action (\ref{eq:e6}) to be stationary for
all $\Phi$ which satisfy (\ref{eq:e0}). By independent variation
of the action (\ref{eq:e6}) with respect to $C$ and $D$, we obtain
\[
\frac{d}{d\tau}\stackrel{>}{C^{A}}=\frac{\partial Tr(\mathcal{H})}{\partial P_{A\dot{E}}}\stackrel{>}{D_{\dot{E}}},\quad\frac{d}{d\tau}\stackrel{<}{D_{A}}=-\stackrel{<\dot{E}}{C\quad}\frac{\partial Tr(\mathcal{H})}{\partial X^{A\dot{E}}}.
\]
 Note that while $\Phi$ does not affect the equations of motion,
it may be necessary to keep it to ensure convergence in the infinite-dimensional
case. Taking the inner product on both sides of these equations with
$\stackrel{<\dot{B}}{C\quad}$ and $\stackrel{>}{D}_{\dot{B}}$ and
applying the Noether constraint (\ref{eq:d7}) and the re\-para\-metri\-zation
(\ref{eq:c17}), we obtain 
\begin{equation}
\frac{dX^{\mu}}{d\overline{\tau}}=\frac{\partial Tr(\mathcal{H})}{\partial P_{\mu}},\quad\frac{dP_{\mu}}{d\overline{\tau}}=-\frac{\partial Tr(\mathcal{H})}{\partial X^{\mu}},\quad\mathcal{H}\equiv\frac{1}{2m}(P^{\mu}P_{\mu}-m^{2}\cdot\underline{1}).\label{eq:d5}
\end{equation}

The dynamical system so obtained describes a general class of unitarily
invariant systems which includes, but is not limited to, systems of
independent particles. Systems of independent particles are obtained
by adding the commutation relations 
\begin{equation}
[X^{\mu},X^{\nu}]=[X^{\mu},P_{\nu}]=0,\label{eq:e7}
\end{equation}
thus allowing all off-diagonal entries in $X$ and $P$, that is all
couplings between different particles, to be gauged away in the same
unitary frame. (\ref{eq:e7}) can be generalized by requiring that
the generators of the Poincaré group be expressed in terms of $X$
and $P$ . This leads to the well-known commutation relations 
\begin{equation}
[X^{\mu},P_{\nu}]=i\hslash\delta_{\nu}^{\mu}\cdot\underline{1},\quad[X^{\mu},X^{\nu}]=0,\quad[P_{\mu},P_{\nu}]=0,\label{eq:e5}
\end{equation}
which are also preserved by the equations of motion. For $\hslash\neq0$,
the couplings between different particles can no longer be gauged
away and the $N$ integral curves of classical dynamics are replaced
with an infinite and irreducible system of coupled paths. The commutation
relations (\ref{eq:e5}) allow $\partial Tr(\mathcal{H})/\partial X$
and $\partial Tr(\mathcal{H})/\partial P$ to be written as commutators,
turning (\ref{eq:d5}) into 
\begin{equation}
\frac{dX^{\mu}}{d\overline{\tau}}=\frac{1}{i\hslash}[X^{\mu},\mathcal{H}],\quad\frac{dP_{\mu}}{d\overline{\tau}}=\frac{1}{i\hslash}[P_{\mu},\mathcal{H}].\label{eq:e23}
\end{equation}
 These equations of motion taken together with the commutation relations
(\ref{eq:e5}), are formally identical to Matrix Mechanics in the
Heisenberg picture and correspond to the Clifford bracket quantization
(\ref{eq:e20})-(\ref{eq:e21}).

In the Clifford space description, the picture independence of quantum
mechanics can be made explicit by coupling an auxiliary unitary gauge
connection to the (vanishing) unitary Noether charge. It transforms
according to 
\[
\Gamma\rightarrow U\Gamma U^{\dagger}-i\frac{dU}{d\tau}U^{\dagger},\quad\bar{\Gamma}(\bar{\tau})=\Gamma(\tau)\frac{d\tau}{d\bar{\tau}},
\]
 and when the ordinary derivatives are replaced by the gauge covariant
derivatives 
\[
\nabla_{\tau}\stackrel{>}{V}\equiv\Bigl(\frac{d}{d\tau}-i\Gamma(\tau)\Bigl)\stackrel{>}{V},\quad\overline{\nabla}_{\bar{\tau}}\stackrel{>}{V}\equiv\Bigl(\frac{d}{d\bar{\tau}}-i\bar{\Gamma}(\bar{\tau})\Bigl)\stackrel{>}{V},
\]
 (\ref{eq:e23}) is turned into 
\[
\overline{\nabla}_{\bar{\tau}}X^{\mu}=\frac{1}{i\hslash}[X^{\mu},\mathcal{H}],\quad\overline{\nabla}_{\bar{\tau}}P_{\mu}=\frac{1}{i\hslash}[P_{\mu},\mathcal{H}].
\]
 The \mbox{Heisenberg} picture corresponds to the gauge $\Gamma=0$.
In the local gauge $\overline{\Gamma}(\overline{\tau})=-\frac{1}{\hslash}\mathcal{H}$,
the commutators on the left and right hand side of (\ref{eq:e23})
cancel out and $X$ and $P$ become stationary. It therefore corresponds
to the Schrödinger picture.

\section{The state vector}

In the foregoing we have seen that when the point particle is described
relative to Clifford space, the classical and the quantum particle
become objects in the same formal system. This makes it possible to
compare directly the classical and the quantum measurement principles.

Let us first consider the classical system $\hslash=0$. In section
4 we found that in a unitary frame where $X$ is diagonal, the paths
in Clifford space constitute a family of integral curves consisting
of eigenvalues of $X$ and $P$. When, for example, the space-time
position $x$ of the particle is being measured at some time $\tau$,
a good measurement would therefore be expected to return an eigenvalue
$x_{i}(\tau)$ of $X(\tau)$. The corresponding Clifford coordinate
$c_{i}(\tau)$ (which for short we shall also call an eigenvalue)
can be expressed as a unitarily invariant expectation value $E$ in
terms of a state vector $|\,s>$: 
\[
c_{i}(\tau)=E(\stackrel{>}{C^{A}})\equiv<s\,|\,\stackrel{>}{C^{A}}.
\]
 To see this, we expand $\stackrel{>}{C}(\tau)$ in terms of $c_{i}(\tau)$:
\[
\stackrel{>}{C^{A}}(\tau)=\sum_{r}|\,x_{r}(\tau)\rangle c_{r}^{A}(\tau),\quad c_{r}^{A}(\tau)\bullet c_{s}^{*\dot{B}}(\tau)=\delta_{rs}x_{s}^{A\dot{B}}(\tau),
\]
 where $|\,x_{i}(\tau)\rangle$ denotes the eigenvectors of $X^{\mu}(\tau)$
with eigenvalues $x_{i}^{\mu}(\tau)$. It follows that the expectation
value $E(\stackrel{>}{C})$ returns the correct value $c_{i}$ of
a measurement when the state vector $|\,s>$ is set equal to the eigenvector
$|\,x_{i}>$. Conversely, if the expectation value coincides with
an eigenvalue, we would expect a good measurement to return this value.
For the purpose of predicting the outcome of future measurements,
the state vector must be subject to a time evolution. In classical
dynamics, we expect that after a measurement has been performed, the
expectation value must stay on the integral curve corresponding to
this measurement. In the unitary gauge where the paths are integral
curves, $X(\tau)$ is diagonal and $\Gamma=0$ . In this frame the
eigenvectors $|\,x_{i}>$ can be chosen to be constants of motion
and the state vector must therefore also be a constant of motion.
This leads to the gauge invariant time evolution 
\begin{equation}
\nabla_{\tau}|s>\equiv(\frac{d}{d\tau}-i\Gamma)\,|s>=0.\label{eq:d13}
\end{equation}

For the classical system, these measurement principles merely represent
a different way of formulating the traditional initial value problem.
Remarkably, however, they also apply to the non-classical system,
regardless of the fact that the way they were derived is no longer
valid. In the quantum system where $X$ and $P$ do not commute, the
assumption that measurements must be expressed through a single state
vector imposes restrictions on which type of measurements can be performed.
The time evolution (\ref{eq:d13}) also holds true, as follows from
the fact that the state vector is known to be a constant of motion
in the \mbox{Heisenberg} picture $\Gamma=0$. The difference between
the classical and the quantum systems becomes clear when we expand
the expectation value $E(C)$ in terms of the eigenvalues $c_{i}$:
\[
E(\stackrel{>}{C^{A}}(\tau))\equiv<s\,|\stackrel{>}{C^{A}}(\tau)=<s\,|\,x_{i}(\tau)>c_{i}(\tau).
\]
 In the classical system, in the gauge $\Gamma=0$, both $<s\,|$
and $|\,x_{i}>$ are constants of motion and hence the expectation
value $E(C(\tau))$ is equal to one of the eigenvalues $c_{i}(\tau)$.
The outcome of a measurement is therefore predictable. This is not
surprising since it was used to derive the time evolution of the state
vector. The role of the state vector in the classical system is simply
to select an integral curve. In the non-classical system, in the gauge
$\Gamma=0$, the state vector is also stationary, but the eigenvectors
$|\,x_{i}>$ undergo a unitary time evolution. After a measurement
has been performed, the expectation value therefore develops into
a complex linear combination of different eigenvalues $c_{i}(\tau)$.
Accordingly, the outcome of a measurement is no longer predictable,
but instead occurs with statistical frequencies given by the Born
rule.

In the non-relativistic limit, the proper time $\overline{\tau}$
is equal to the expectation value of $X^{0}$ which represents the
`physical' time $t\equiv<s|X^{0}|s>$: 
\[
\frac{dt}{d\overline{\tau}}=<s|\bar{\nabla}_{\overline{\tau}}X^{0}|s>=\frac{1}{m}<s|P^{0}|s>\approx1,
\]
 where we have used the time evolution of the state vector and the
equations of motion for $X^{0}$. Restricting the equations of motion
(\ref{eq:e23}) to $\mu=1,2,3\:$, the Hamiltonian density $\mathcal{H}$
effectively reduces to the non-relativistic Hamiltonian density 
\[
\tilde{\mathcal{H}}=\frac{1}{2m}(P_{x}^{2}+P_{y}^{2}+P_{z}^{2}).
\]
 Taken together with the corresponding commutation relations, this
system is identical to that of non-relativistic Matrix Mechanics.
The Schrödinger picture corresponds to the non-relativistic gauge
condition $\overline{\Gamma}(t)=-\frac{1}{\hslash}\tilde{\mathcal{H}}$
which turns the time evolution (\ref{eq:d13}) of the state vector
into the matrix form of the Schrödinger equation.

\section{The classical string in Clifford space}

The world sheet of the relativistic string in Clifford space is described
by the coordinate functions $c^{A}(\tau,\sigma)$ with values in the
generating space of the infinite-dimensional Clifford algebra obtained
from (\ref{eq:d20}) by letting $N\rightarrow\infty$. We follow the
convention that $\mu,\nu,\ldots$ denote the space-time indices and
$\alpha,\beta,\ldots$, the world sheet indices. Differentiation with
respect to the world sheet parameters $\sigma^{\alpha}=\tau,\sigma$
will be written as $\partial_{\alpha}$.

It is well known that for a string which resides in space-time, the
Lorentz metric $\eta_{\mu\nu}$ induces a metric on the world sheet
through the tangent derivatives $\partial_{\alpha}x^{\mu}$. For a
string which resides in Clifford space, we use the complex vectors
\[
V_{\alpha}^{\mu}\equiv\sigma_{A\dot{B}}^{\mu}c^{A}\bullet\partial_{\alpha}c^{*\dot{B}},
\]
 which have the real part $\partial_{\alpha}x^{\mu}$. These vectors
induce the Hermitian tensor 
\[
g_{\alpha\beta}\equiv V_{\alpha}^{\mu}V_{\beta}^{\nu*}\eta_{\mu\nu},\quad g_{\alpha\beta}^{*}=g_{\beta\alpha},
\]
 on the Clifford worldsheet, which can be decomposed into a real symmetric
tensor $h_{\alpha\beta}$ and a real scalar $\phi$: 
\[
g_{\alpha\beta}=h_{\alpha\beta}+i\phi\sqrt{h}\,\epsilon_{\alpha\beta},\quad h_{\alpha\beta}\equiv g_{(\alpha\beta)},\quad\phi\equiv-\frac{1}{2}ih^{-\frac{1}{2}}\epsilon^{\alpha\beta}g_{\alpha\beta},\quad h\equiv|det(h_{\alpha\beta})|.
\]

The re\-para\-metri\-zation invariant string generalization of
the point particle action (\ref{eq:c21}) is 
\begin{equation}
I=\int d\tau d\sigma\bigl(3\sqrt[3]{W^{\mu}W_{\mu}}-m^{2}\bigl)\sqrt{h}\,,\quad W^{\mu}\equiv\frac{1}{2}\sigma_{A\dot{B}}^{\mu}h^{\alpha\beta}\partial_{\alpha}c^{A}\bullet\partial_{\beta}c^{*\dot{B}}.\label{eq:f14}
\end{equation}
 To write this action in an explicit covariant first order form, we
use De Donder-Weyl covariant canonical variables \cite{key-2,key-15,key-11}.
The polymomenta density conjugate to $c$ is 
\[
d_{A}^{*\alpha}\equiv\sqrt{h}\mathrm{d}_{A}^{^{*}\alpha}\equiv\frac{\partial L}{\partial(\partial_{\alpha}c^{A})}=(W^{\nu}W_{\nu})^{-\frac{2}{3}}W_{\mu}\sigma_{A\dot{B}}^{\mu}h^{\alpha\beta}\partial_{\beta}c^{*\dot{B}}\sqrt{h}\,,
\]
 where $L$ is the Lagrangian density in (\ref{eq:f14}). This leads
to the expressions 
\begin{multline*}
\frac{1}{2}h_{\alpha\beta}\,\mathrm{d}^{*\alpha A}\bullet\mathrm{d}^{\dot{\beta B}}\,h_{\gamma\delta}\mathrm{d}_{A}^{*\gamma}\bullet\mathrm{d}_{\dot{B}}^{\delta}=\sqrt[3]{W^{\mu}W_{\mu}}\,,\quad\mathrm{d}_{A}^{*\alpha}\bullet\partial_{\alpha}c^{A}+c.c.=4\sqrt[3]{W^{\mu}W_{\mu}}\,,
\end{multline*}
 from which we obtain the De Donder-Weyl scalar Hamiltonian density
\begin{multline*}
\mathcal{H}\equiv d_{A}^{*\alpha}\bullet\partial_{\alpha}c^{A}+c.c.-L=\sqrt{h}\,(p^{\mu}p_{\mu}+m^{2}),\quad p^{\mu}\equiv\frac{1}{2}\sigma_{A\dot{B}}^{\mu}h_{\alpha\beta}\,\mathrm{d}^{*\alpha A}\bullet\mathrm{d}^{\beta\dot{B}},
\end{multline*}
 and hence the first order form 
\[
I_{M}=\int d\tau d\sigma L_{M},\quad L_{M}=\sqrt{h}\,\Bigl(\mathrm{d}_{A}^{*\alpha}\bullet\partial_{\alpha}c^{A}+c.c.-(p^{\mu}p_{\mu}+m^{2})\Bigr)
\]
 of the Polyakov action (\ref{eq:f14}). Without kinetic terms for
$h_{\alpha\beta}$ and $\phi$, the equations of motion for the metric
would lead to a singular metric. The simplest such kinetic term, is
a Jackiw-Teitelboim term \cite{key-12,key-13} which describes a world
sheet with constant curvature 
\[
I_{JT}=\int d\tau d\sigma L_{JT},\quad L_{JT}=\sqrt{h}\,\Bigl(\phi(R(h_{\alpha\beta})-2\Lambda)\Bigr).
\]
 The equations of motion obtained from $L=L_{M}+L_{JT}$ are 
\begin{equation}
\partial_{\alpha}c^{A}=h_{\alpha\beta}\,p^{A\dot{E}}\mathrm{d}_{\dot{E}}^{\beta},\label{eq:f51}
\end{equation}
 
\begin{equation}
\partial_{\alpha}(d_{A}^{*\alpha})=0,\label{eq:f52}
\end{equation}
 
\[
\frac{1}{\sqrt{h}}\frac{\partial L}{\partial\phi}=R(h_{\alpha\beta})-2\Lambda=0,
\]
 
\[
\frac{1}{\sqrt{h}}\frac{\partial L}{\partial h_{\alpha\beta}}=-(h^{\alpha\beta}\nabla^{2}+\Lambda h^{\alpha\beta}-\nabla^{\alpha}\nabla^{\beta})\phi+T^{\alpha\beta}=0,
\]

\begin{multline}
T^{\alpha\beta}\equiv\frac{1}{\sqrt{h}}\frac{\partial L_{M}}{\partial h_{\alpha\beta}}=\frac{1}{2}\bigl(\mathrm{d}_{A}^{*\gamma}\bullet\partial_{\gamma}c^{A}+c.c.-p_{\mu}p^{\mu}-m^{2}\bigr)h^{\alpha\beta}-p^{\mu}\sigma_{\mu}^{A\dot{B}}\mathrm{d}_{A}^{*(\alpha}\bullet\mathrm{d_{\dot{B}}^{\beta)}}\\
=\frac{1}{2}(3p_{\mu}p^{\mu}-m^{2})h^{\alpha\beta}-p^{\mu}\sigma_{\mu}^{A\dot{B}}\mathrm{d}_{A}^{*(\alpha}\bullet\mathrm{d}_{\dot{B}}^{\beta)},\label{eq:f72}
\end{multline}
 where $\nabla_{\alpha}$ denotes the covariant derivative.

To find the wave solutions, we consider a flat world sheet ($\Lambda=0$)
and the parametrization $h_{\alpha\beta}=\eta_{\alpha\beta}.$ Traveling
waves with a spatial period of $4\pi$ are described by 
\begin{equation}
c^{A}=k^{A}+l^{A}\tau+\sum_{n\neq0}a_{n}^{A}e^{i\frac{1}{2}n(\tau+\sigma)}+\sum_{n\neq0}b_{n}^{A}e^{i\frac{1}{2}n(\tau-\sigma)},\quad0\leq\sigma\leq\pi.\label{eq:f73}
\end{equation}
 With the exception of $a_{n}\bullet a_{-n}^{*}$ and $b_{n}\bullet b_{-n}^{*}$,
we assume that the inner products between different coefficients vanish.
This leads to the space-time trajectories 
\begin{multline}
x^{A\dot{B}}=k^{A}\bullet k^{*\dot{B}}+l^{A}\bullet l^{*\dot{B}}\tau^{2}\\
+\sum_{n\neq0}\Bigl(a_{n}^{A}\bullet a_{n}^{*\dot{B}}+b_{n}^{A}\bullet b_{n}^{*\dot{B}}+a_{n}^{A}\bullet a_{-n}^{*\dot{B}}e^{in(\tau+\sigma)}+b_{n}^{A}\bullet b_{-n}^{*\dot{B}}e^{in(\tau-\sigma)}\Bigr),\label{eq:f74}
\end{multline}
 with a spatial period of $2\pi$. When $p^{2}\equiv p_{\mu}p^{\mu}\neq0$,
(\ref{eq:f51}) can be solved with respect to $p^{A\dot{B}}$ and
$\mathrm{d}_{\dot{E}}^{\beta}$, giving 
\[
p^{2}p^{A\dot{B}}=\eta^{\alpha\beta}\partial_{\alpha}c^{A}\bullet\partial_{\beta}c^{*\dot{B}}=l^{A}\bullet l^{*\dot{B}},\quad\mathrm{d}_{\beta\dot{E}}=(p^{2})^{-2}l_{\dot{E}}^{*}\bullet l_{A}\partial_{\beta}c^{A}.
\]
 It follows that $p_{\mu}$ is constant and, since $c^{A}$ satisfies
the free wave equation, the polymomenta $\mathrm{d}_{\dot{A}}^{\beta}$
must satisfy their equation of motion (\ref{eq:f52}). The space-time
coordinates (\ref{eq:f74}) satisfy $\square x^{A\dot{B}}=2l^{A}\bullet l^{*\dot{B}}$
and therefore the Clifford string is not a substructure of the bosonic
string. In a time interval of given length, it approaches the bosonic
string in the limit $l^{A}\rightarrow0$ with $l^{A}\bullet l^{*\dot{B}}\tau$
held fixed.

As shown by Navarro \cite{key-16}, the equations of motion for $h_{\alpha\beta}$
and $\phi$ are equivalent to the single equation 
\begin{equation}
\nabla_{\alpha}\nabla_{\beta}\phi=-h_{\alpha\beta}\Lambda\phi+h_{\alpha\beta}T-T_{\alpha\beta}.\label{eq:f69}
\end{equation}
 Since $p_{\mu}$ is constant, we can set $p_{\mu}p^{\mu}=m^{2}$.
This makes the trace $T$ of the energy-momentum tensor (\ref{eq:f72})
vanish and (\ref{eq:f69}) reduce to 
\begin{equation}
\partial_{\alpha}\partial_{\beta}\phi=-m^{2}\eta_{\alpha\beta}+\mathrm{\eta_{\gamma\delta}d_{A}^{*\gamma}\bullet\mathrm{d_{\dot{B}}^{\delta}}}\mathrm{d}_{(\alpha}^{*A}\bullet\mathrm{d_{\beta)}^{\dot{B}}}.\label{eq:f90}
\end{equation}
 The trace of the right hand side vanishes and $\phi$ therefore satisfies
the free wave equation with the general solution $\phi=\phi_{L}(\tau+\sigma)+\phi_{R}(\tau-\sigma)$.
Reinserting this solution into (\ref{eq:f90}) gives the differential
equation for $\phi_{L}$ 
\begin{equation}
\phi_{L}''(\tau+\sigma)=\frac{1}{2}m^{2}-\frac{1}{4}m^{-2}l_{A}\bullet l_{\dot{B}}^{*}\sum_{n\neq0}n^{2}\bigl(a_{n}^{A}\bullet a_{n}^{*\dot{B}}-a_{n}^{A}\bullet a_{-n}^{*\dot{B}}e^{in(\tau+\sigma)}\bigr),\label{eq:f70}
\end{equation}
 and the corresponding one for $\phi_{R}$ obtained from (\ref{eq:f70})
by $a\rightarrow b$ and $\sigma\rightarrow-\sigma$. Integrating
these equations yields the dilaton field 
\begin{multline*}
\phi=k+k_{\alpha}\sigma^{\alpha}+\frac{1}{2}m^{2}(\tau^{2}+\sigma^{2})-\frac{1}{4}m^{-2}l_{A}\bullet l_{\dot{B}}^{*}\sum_{n\neq0}\Bigl(\frac{1}{2}n^{2}a_{n}^{A}\bullet a_{n}^{*\dot{B}}(\tau+\sigma)^{2}\\
+\frac{1}{2}n^{2}b_{n}^{A}\bullet b_{n}^{*\dot{B}}(\tau-\sigma)^{2}+a_{n}^{A}\bullet a_{-n}^{*\dot{B}}e^{in(\tau+\sigma)}+b_{n}^{A}\bullet b_{-n}^{*\dot{B}}e^{in(\tau-\sigma)}\Bigr).
\end{multline*}

When $\Gamma$ is a space-like curve connecting two points on the
boundaries of the world sheet, the total Clifford momentum $\mathrm{d}_{A}^{*tot}$
of the string can be determined by integration of the conserved current
density $d_{A}^{*\alpha}$ 
\begin{equation}
\mathrm{d}_{A}^{*tot}\equiv\int_{\Gamma}d\sigma^{\alpha}\epsilon_{\beta\alpha}d_{A}^{*\beta}.\label{eq:f80}
\end{equation}
 It is reasonable to assume that the total Clifford momentum determines
the total space-time momentum of the string in the same way as for
the point particle. For $a_{n}=b_{n}=0$, we get $d_{A}^{*tot}\bullet d_{\dot{B}}^{tot}=\pi^{2}p_{A\dot{B}}$
which identifies $p_{\mu}$ and $m$ as the space-time momentum and
mass of the non-vibrating string of unit length.

The spinning string is described by a subset of the trajectories (\ref{eq:f73})
\[
c^{1}=k\tau+a\,e^{i\frac{1}{2}(\tau+\sigma)}+b\,e^{i\frac{1}{2}(\tau-\sigma)},\;c^{2}=l\tau+a\,e^{-i\frac{1}{2}(\tau+\sigma)}+b\,e^{-i\frac{1}{2}(\tau-\sigma)},
\]
 where $a\bullet a^{*}=b\bullet b^{*}$, $k\bullet k^{*}=l\bullet l^{*}$
and all other inner products vanish. This produces the space-time
trajectories 
\begin{multline*}
x=\frac{1}{2}(c^{1}\bullet c^{*\dot{2}}+c^{*\dot{1}}\bullet c^{2})=2a\bullet a^{*}\cos(\tau)\cos(\sigma),\;y=\frac{1}{2i}(c^{1}\bullet c^{*\dot{2}}-c^{*\dot{1}}\bullet c^{2})\\
=2a\bullet a^{*}\sin(\tau)\cos(\sigma),\;z=\frac{1}{2}(c^{1}\bullet c^{*\dot{1}}-c^{2}\bullet c^{*\dot{2}})=0,\\
t=\frac{1}{2}(c^{1}\bullet c^{*\dot{1}}+c^{2}\bullet c^{*\dot{2}})=2a\bullet a^{*}+k\bullet k^{*}\tau^{2}.
\end{multline*}

\section{The quantum string}

If the quantized string is going to represent a physical particle,
the Lorentz charge and space-time momentum of the string must satisfy
the Poincaré algebra. The conserved $SL(2.\mathbb{C})$ and $U(1)$
Noether current densities are 
\[
\jmath_{AB}^{\alpha}\equiv(c_{A}\bullet d_{B}^{*\alpha}+c_{B}\bullet d_{A}^{*\alpha}),\quad i^{\alpha}\equiv i(c^{A}\bullet d_{A}^{*\alpha}-c.c.).
\]
 Let $\Gamma$ be a space-like curve connecting two fixed points on
the boundaries of the world sheet and let $\sigma^{\alpha}(u)$, $\sigma^{\alpha}(u')$
and $\sigma^{\alpha}(u'')$ be three points on this curve. Then we
can define the world sheet scalars 
\[
\mathrm{j}_{AB}\equiv v^{\alpha}\epsilon_{\beta\alpha}\jmath_{AB}^{\beta}=c_{A}\bullet\mathrm{d_{B}^{*}}+c_{B}\bullet\mathrm{d_{A}^{*}},\quad\mathrm{d_{A}^{*}}\equiv v^{\alpha}\epsilon_{\beta\alpha}d_{A}^{*\beta},\quad v^{\alpha}\equiv\frac{d\sigma^{\alpha}}{du},
\]
 and the total $SL(2.\mathbb{C})$ charge 
\[
\mathrm{j}_{AB}^{tot}\equiv\int_{\Gamma}d\sigma^{\alpha}\epsilon_{\beta\alpha}\jmath_{AB}^{\beta}=\int du\,\mathrm{j_{AB}},
\]
 which is path independent. The Clifford bracket is defined as
\begin{multline*}
\Bigl\{\mathrm{j}_{AB}^{'},\mathrm{j}_{EF}^{''}\Bigr\}_{C.B.}\equiv\int du\frac{1}{2}\Bigl(\biggl\{\frac{\partial\mathrm{j}_{AB}^{'}}{\partial c^{G}},\frac{\partial\mathrm{j}_{EF}^{''}}{\partial\mathrm{d}_{G}^{*}}\biggr\}+\biggl\{\frac{\partial\mathrm{j}_{AB}^{'}}{\partial c^{*\dot{G}}},\frac{\partial\mathrm{j}_{EF}^{''}}{\partial\mathrm{d}_{\dot{G}}}\biggr\}\\
-\biggl\{\frac{\partial\mathrm{j}_{EF}^{''}}{\partial c^{G}},\frac{\partial\mathrm{j}_{AB}^{'}}{\partial\mathrm{d}_{G}^{*}}\biggr\}-\biggl\{\frac{\partial\mathrm{j}_{EF}^{''}}{\partial c^{*\dot{G}}},\frac{\partial\mathrm{j}_{AB}^{'}}{\partial\mathrm{d}_{\dot{G}}}\biggr\}\Bigr),
\end{multline*}
where unprimed variables depend on $u$, and variables with a single
prime or a double prime depend on $u'$ and $u''$ respectively. By
means of the functional differentiation rule $\partial f^{A}(u')/\partial f^{B}(u)=\delta_{B}^{A}\delta(u'-u)$,
the Clifford brackets reduce to
\begin{multline}
\left\{ \mathrm{j}_{AB}^{'},\mathrm{j}_{EF}^{''}\right\} _{C.B.}=\Bigl((\mathrm{j}_{AE}^{'}\,\epsilon_{FB}+A\leftrightarrow B)+E\leftrightarrow F\Bigr)\delta(u'-u''),\\
\Bigl\{\mathrm{j}_{AB}^{'},\mathrm{j}_{\dot{E}\dot{F}}^{''}\Bigr\}_{C.B.}=0,\label{eq:g1}
\end{multline}
which is turned into the commutation relations
\begin{multline}
[J_{AB}^{'},J_{EF}^{''}]=i\hslash\Bigl((J_{AE}^{'}\,\epsilon_{FB}+A\leftrightarrow B)+E\leftrightarrow F\Bigr)\delta(u'-u''),\quad[J_{AB}^{'},J_{\dot{E}\dot{F}}^{\dagger''}]=0,\\
J_{AB}\equiv v^{\alpha}\epsilon_{\beta\alpha}J_{AB}^{\beta}=\stackrel{>}{C_{A}}\bullet\stackrel{<}{D_{B}}+\stackrel{>}{C_{B}}\bullet\stackrel{<}{D_{A}},\quad\stackrel{<}{D_{A}}\equiv v^{\alpha}\epsilon_{\beta\alpha}\stackrel{<\beta}{D_{A}}\label{eq:g2}
\end{multline}
by the quantization $c\rightarrow\stackrel{>}{C}$, $d^{*\alpha}\rightarrow\stackrel{<\alpha}{D\quad}$,
$\{,\}_{C.B.}\rightarrow\frac{1}{i\hslash}[,]$. Since $v^{\alpha}$
is a space-like vector and its weight is different from 1, there exists
a parametrization in which $v^{\alpha}=(0,1)$ and where (\ref{eq:g2})
becomes the equal-time commutation relations
\begin{multline*}
[J_{AB}^{1}(\tau,\sigma),J_{EF}^{1}(\tau,\sigma')]=i\hslash\Bigl((J_{AE}^{1}(\tau,\sigma)\,\epsilon_{FB}+A\leftrightarrow B)+E\leftrightarrow F\Bigr)\delta(\sigma-\sigma'),\\
{}[J_{AB}^{1}(\tau,\sigma),J_{\dot{E}\dot{F}}^{\dagger1}(\tau,\sigma')]=0.
\end{multline*}
Integrating both sides of (\ref{eq:g2}) with respect to $u'$ and
$u''$, we obtain the path-independent algebra
\begin{multline}
[J_{AB}^{tot},J_{EF}^{tot}]=i\hslash\Bigl((J_{AE}^{tot}\,\epsilon_{FB}+A\leftrightarrow B)+E\leftrightarrow F\Bigr),\quad[J_{AB}^{tot},J_{\dot{E}\dot{F}}^{\dagger tot}]=0,\\
J_{AB}^{tot}\equiv\int_{\Gamma}d\sigma^{\alpha}\epsilon_{\beta\alpha}J_{AB}^{\beta},\quad J_{AB}^{\beta}\equiv\stackrel{>}{C_{A}}\bullet\stackrel{<\beta}{D_{B}}+\stackrel{>}{C_{B}}\bullet\stackrel{<\beta}{D_{A}}.\label{eq:g3}
\end{multline}
The symmetric second rank spinor $J_{AB}^{tot}$ is equivalent to
the skew-symmetric tensor
\[
M_{\mu\nu}\equiv\frac{1}{4}\sigma_{\mu}^{A\dot{E}}\sigma_{\nu}^{B\dot{F}}(J_{AB}^{tot}\epsilon_{\dot{E}\dot{F}}+J_{\dot{E}\dot{F}}^{\dagger tot}\epsilon_{AB}),\quad J_{AB}^{tot}=\frac{1}{2}\epsilon^{\dot{E}\dot{F}}\sigma_{A\dot{E}}^{\mu}\sigma_{B\dot{F}}^{\nu}M_{\mu\nu},
\]
with hermitian components $M_{\mu\nu}^{\dagger}=M_{\mu\nu}$, and
when written in terms of $M_{\mu\nu}$ the algebra (\ref{eq:g3})
is seen to be the Lorentz algebra. 

The usual proof \cite{key-18} that orbital angular momentum may only
take on integral values of $\hslash$ does not apply to $J_{AB}$.
Clifford strings with half-integral spin could provide a more detailed
picture of a fermion than is possible in a space-time description.

The foregoing procedure for obtaining path-independent commutation
relations applies to conserved current densities $N^{\alpha}$ and
$M^{\alpha}$ for which $v^{\alpha}\epsilon_{\beta\alpha}N^{\beta}$
and $v^{\alpha}\epsilon_{\beta\alpha}M^{\beta}$ depend on $c^{A}$
and $\mathrm{d_{A}^{*}}\equiv v^{\alpha}\epsilon_{\beta\alpha}d_{A}^{*\beta}$
only. To include the space-time momentum in the commutation relations,
we need a conserved current density which closes with the $SL(2.\mathbb{C})$
current under Clifford brackets, and which has a total charge that
can be identified with the total space-time momentum of the string.
These two conditions single out the conserved current density
\[
p_{A\dot{B}}^{\alpha}\equiv\mathrm{d}_{\dot{B}}^{tot}\bullet d_{A}^{*\alpha}.
\]
This is not a hermitian spinor and therefore cannot be interpreted
as a local space-time momentum current. It does, however, have the
correct total charge $\mathrm{p}_{A\dot{B}}^{tot}=\mathrm{d}_{\dot{B}}^{tot}\bullet\mathrm{d}_{A}^{*tot}$
and closes with the $SL(2.\mathbb{C})$ current
\begin{multline*}
\{\mathrm{j}_{AB}^{''},\mathrm{p}_{E\dot{F}}^{'}\}_{C.B.}=(\epsilon_{AE}\mathrm{p}_{B\dot{F}}^{'}+\epsilon_{BE}\mathrm{p}_{A\dot{F}}^{'})\delta(u'-u''),\quad\{\mathrm{p}_{A\dot{B}}^{'},\mathrm{p}_{E\dot{F}}^{''}\}_{C.B.}=0,\\
\mathrm{p}_{A\dot{B}}\equiv v^{\alpha}\epsilon_{\beta\alpha}p_{A\dot{B}}^{\alpha}.
\end{multline*}
These brackets are turned into the commutation relations
\begin{multline}
[J_{AB}^{''},P_{E\dot{F}}^{'}]=i\hslash(\epsilon_{AE}P_{B\dot{F}}^{'}+\epsilon_{BE}P_{A\dot{F}}^{'})\delta(u'-u''),\quad[P_{A\dot{B}}^{'},P_{E\dot{F}}^{''}]=0,\\
P_{A\dot{B}}\equiv\stackrel{>}{D_{\dot{B}}^{tot}}\bullet\stackrel{<}{D_{A}},\quad\stackrel{<}{D_{A}^{tot}}\equiv\int_{\Gamma}d\sigma^{\alpha}\epsilon_{\beta\alpha}\stackrel{<\beta}{D_{A}}\label{eq:g4}
\end{multline}
by the quantization $c\rightarrow\stackrel{>}{C}$, $d^{*\alpha}\rightarrow\stackrel{<\alpha}{D\quad}$,
$\{,\}_{C.B.}\rightarrow\frac{1}{i\hslash}[,]$. Integrating both
sides of (\ref{eq:g4}) with respect to $u'$ and $u''$, we obtain
the path-independent commutation relations
\begin{multline}
[J_{AB}^{tot},P_{E\dot{F}}^{tot}]=i\hslash(\epsilon_{AE}P_{B\dot{F}}^{tot}+\epsilon_{BE}P_{A\dot{F}}^{tot}),\quad[P_{A\dot{B}}^{tot},P_{E\dot{F}}^{tot}]=0,\\
P_{A\dot{B}}^{tot}\equiv\stackrel{>}{D_{\dot{B}}^{tot}}\bullet\stackrel{<}{D_{A}^{tot}.}\label{eq:g6}
\end{multline}
When written in terms of $M_{\mu\nu}$ and $P_{\mu}^{tot}$, the algebra
(\ref{eq:g3}),(\ref{eq:g6}) is seen to be the Poincaré algebra. 

\section{Conclusion}

We have described the dynamics of the classical point particle in
terms of the Clifford substructure of its canonical variables and
shown that this substructure itself forms a canonical system. Compared
to the space-time description, this description offers a conceptually
simpler road to matrix mechanics because the Clifford algebra inherently
supports the unitary symmetry of quantum states through its generating
algebra. Unlike the point particle, we found that the relativistic
string in Clifford space is not a substructure of the bosonic string
in space-time. We found the wave solutions for a flat world sheet
and showed that the $SL(2.\mathbb{C})$ charge and space-time momentum
of the quantized string satisfy the Poincaré algebra.

There are good reasons to believe that a four-dimensional Minkowski
space does not suffice to accommodate the particle physics of the
Standard Model. The \mbox{Clifford} model discussed in the foregoing
is limited to a four-dimensional Minkowski space because it is based
on complex Weyl spinors. Since Weyl spinors are an integral part of
the model, it is difficult to see how the dimension of space-time
can be increased without replacing the complex numbers with a higher
dimensional algebra. The complex numbers correspond to the \mbox{Clifford}
algebra $\mathcal{C}l(0,1,\mathbb{R})$. Increasing the dimension,
we find $\mathcal{C}l(0,2,\mathbb{R})$ which corresponds to the quaternions
and $\mathcal{C}l(0,3,\mathbb{R})$ which can be deformed into the
octonions. For algebraic reasons \cite{key-8,key-9,key-10}, such
spinors would be expected to generate a six-dimensional and a ten-dimensional
Minkowski space respectively.

\end{document}